\documentclass[aps,twocolumn,english,aip,superscriptaddress,amssymb, showpacs,preprintnumbers,amsmath,amssymb,superscriptaddress]{revtex4-1}

\usepackage{graphicx}% Include figure files
\usepackage{dcolumn}% Align table columns on decimal point
\usepackage{bm}% bold math
\newcommand{\nc}{\newcommand}
\nc{\be}{\begin{equation}}
\nc{\ee}{\end{equation}}
\nc{\bea}{\begin{eqnarray}}
\nc{\eea}{\end{eqnarray}}
\nc{\bean}{\begin{eqnarray*}}
\nc{\eean}{\end{eqnarray*}}
\nc{\mb}{\mbox}
\nc{\rnc}{\renewcommand}
\nc{\vk}{\mb{\bf k}}
\nc{\vp}{\mb{\bf p}}
\nc{\vn}{\mb{\bf n}}
\nc{\vq}{\mb{\bf q}}
\nc{\rr}{\mb{\bf r}}
\nc{\vz}{\hat {\mb{\bf z}}}
\nc{\vj}{\mb{\boldmath$j$}}
\nc{\vg}{\mb{\boldmath$g$}}
\nc{\x}{\mb{\boldmath$x$}}
\nc{\A}{\mb{\boldmath$A$}}
\nc{\va}{\mb{\boldmath$a$}}
\nc{\vs}{\mb{\boldmath$\sigma$}}
\nc{\vpi}{\mb{\boldmath$\pi$}}
\nc{\nab}{\nabla}
\nc{\X}{\sf x}

\begin{document}
\title{Three-dimensional dynamics of magnetic hopfion driven by spin transfer torque}

\author{Yizhou Liu}
\affiliation{Beijing National Laboratory for Condensed Matter Physics, Institute of Physics, Chinese Academy of Sciences, Beijing 100190, China}

\author{Wentao Hou}
\affiliation{Department of Physics and Astronomy, University of New Hampshire, Durham, New Hampshire 03824, USA}

\author{Xiufeng Han}
\thanks{xfhan@iphy.ac.cn}
\affiliation{Beijing National Laboratory for Condensed Matter Physics, Institute of Physics, Chinese Academy of Sciences, Beijing 100190, China}
\affiliation{Center of Materials Science and Optoelectronics Engineering, University of Chinese Academy of Sciences, Beijing 100049, China}
\affiliation{Songshan Lake Materials Laboratory, Dongguan, Guangdong 523808, China}

\author{Jiadong Zang}
\thanks{Jiadong.Zang@unh.edu}
\affiliation{Department of Physics and Astronomy, University of New Hampshire, Durham, New Hampshire 03824, USA}
\affiliation{Materials Science Program, University of New Hampshire, Durham, New Hampshire 03824, USA}

\begin{abstract}

%Magnetic hopfions, three-dimensional (3D) topological solitons, host complicated spin configurations.
Magnetic hopfion is three-dimensional (3D) topological soliton with novel spin structure that would enable exotic dynamics.
Here we study the current driven 3D dynamics of a magnetic hopfion with unit Hopf index in a frustrated magnet.
Attributed to spin Berry phase and symmetry of the hopfion, the phase space entangles multiple collective coordinates, thus the hopfion exhibits rich dynamics including longitudinal motion along the current direction, transverse motion perpendicular to the current direction, rotational motion and dilation.
Furthermore, the characteristics of hopfion dynamics is determined by the ratio between the non-adiabatic spin transfer torque parameter and the damping parameter. 
Such peculiar 3D dynamics of magnetic hopfion could shed light on understanding the universal physics of hopfions in different systems and boost the prosperous development of 3D spintronics.

\end{abstract}


\maketitle

{\em Introduction}---Hopfions are three-dimensional (3D) topological solitons initially proposed in the Skyrme-Faddeev model~\cite{skyrme_nonlinear_1961, skyrme_unified_1962, faddeev_comments_1976}.
The three spatial dimensions endow hopfions with diverse configurations such as rings, links, and knots that can be classified by the Hopf index $Q_H$, a topological index that characterizes the homotopy group $\Pi_3(S^2)$ classifying maps from $S^3$ to $S^2$~\cite{hopf_uber_1931, faddeev_stable_1997, battye_knots_1998, hietarinta_faddeev_hopf_1999}.
Although hopfions were first studied in the contents of field theories, they turn out to emerge in various physical systems, such as optics, liquid crystals, Bose-Einstein condensates, superconductors, etc~\cite{dennis_isolated_2010, ackerman_static_2017, volovik_1977, babaev_hidden_2002, hall_tying_2016, kawaguchi_knots_2008, babaev_dual_2002}.  
Very recently, their magnetic counterparts have been theoretically proposed in frustrated magnets~\cite{sutcliffe_skyrmion_2017, rybakov_magnetic_2019} and confined chiral magnetic heterostructures~\cite{liu_binding_2018, sutcliffe_hopfions_2018, tai_static_2018}, further stimulating the study of hopfion from a new respect.

\begin{figure}
%\begin{center}
\includegraphics[width=3.2in,height=2.7in]{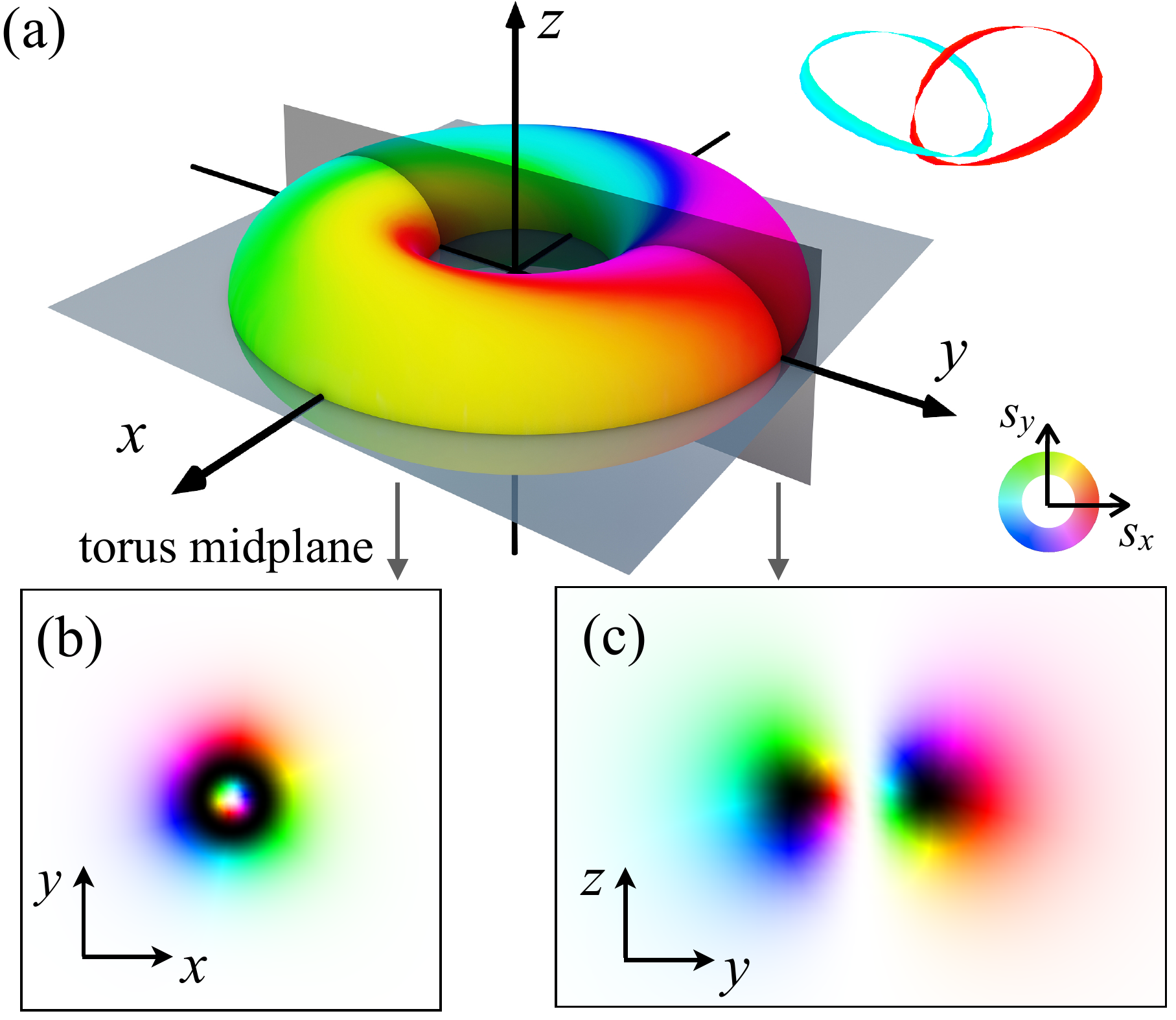}
\caption{(a) Iso-spin contours with $S_z=0$ for a magnetic hopfion with $Q_H=1$. 
Inset is the iso-spin contours of ${\bf S}= + \hat{x}$ (red) and ${\bf S}= - \hat{x}$ (cyan) that demonstrate the unity linking number of the hopfion.
(b) and (c) are the cross-sections of hopfion onto $xy$ (b) and $yz$ (c) planes, as depicted by the grey arrows.
At the initial state, the torus midplane lies in the $xy$ plane.
In the color scheme, black indicates $S_z=-1$ and white indicates $S_z=1$. 
The color wheel stands for in-plane spin directions.
}
\label{fig:schematic}
%\end{center}
\end{figure}

While the sophisticated configurations of hopfion could give rise to fascinating physical phenomena~\cite{faddeev_stable_1997, radu_stationary_2008}, many of their physical properties, especially their dynamics, are still largely unexplored. 
Low-dimensional magnetic topological solitons like skyrmions and vortices have been extensively studied over the past few decades with long lasting interests in both their fundamental physical properties and potential applications~\cite{nagaosa_topological_2013, fert_skyrmions_2013, Guslienko_vortex_2008}. 
%
%Many different methods have been developed to excite the dynamics of topological magnetic solitons.
%
Therefore, it is also important to unravel the dynamics of the magnetic hopfion, especially its most essential dynamics driven by the spin transfer torque (STT) under electric current.
Hopfion dynamics have been recently studied in confined chiral magnetic heterostructures~\cite{wang_current_driven_2019}.
But in this case, hopfions are only allowed to move in two spatial dimensions and the unique physics associated with the third spatial dimension is completely suppressed by the strong boundary condition.
In this Letter, we investigate the current-induced dynamics of a magnetic hopfion in frustrated magnet, where hopfions are free to move in all directions and their full 3D dynamics can be explored.
The hopfion studied here has $Q_H = 1$ and its motion is driven by both the adiabatic and non-adiabatic STT effect~\cite{slonczewski_current_driven_1996, berger_emission_1996, zhang_roles_2004}.
Based on the symmetry of hopfion's spin configuration (Fig.~\ref{fig:schematic}), two typical cases are studied, i.e., current in the torus midplane and current perpendicular to the torus midplane.
As manifested by its 3D configuration, hopfion possesses various types of dynamics including translational motion, rotation, and dilation. 
The spin Berry phase of hopfion hosts an entangled phase space, which further conjugates these dynamics and gives rise to more exotic dynamical properties.
All these dynamical behavior can be captured by an analytical model derived in terms of multi-dimensional collective coordinates and generalized Thiele's approach.
A phenomenological analysis is also employed to bridge the dynamics of hopfion and skyrmion string.

{\em Spin Berry phase and entangled phase space}---
%
%
%
%The preimage, i.e., the iso-spin surface that corresponds to a Hopf map of a point on the $S^2$ unit sphere to 3D space, 
%
%
We consider here a hopfion with $Q_H = 1$.
A typical hopfion configuration can be achieved by a stereographic projection from $\mathbb{R}^3$ to $S^3$: ${\mathbf \chi}=((x/r)\sin f, (y/r)\sin f, (z/r)\sin f, \cos f)$, followed by the Hopf map ${\mathbf S_0}=\langle z|{\mathbf \sigma}|z\rangle$, where the spinor $|z\rangle=(\chi_4+i\chi_3, \chi_1+i\chi_2)^T$, $r^2 = x^2 + y^2 + z^2$, and $f$ is a function of $r$ satisfying the boundary conditions $f(0)=\pi$ and $f(\infty)=0$~\cite{hietarinta_faddeev_hopf_1999}.
Explicitly the configuration is given by:
\begin{equation} 
\begin{aligned}
\label{Eq:ansatz}
S_{0}^x &= \frac{x}{r}\text{sin}2f +  \frac{yz}{r^2}\text{sin}^2 f,\\
S_{0}^y &= \frac{y}{r}\text{sin}2f -  \frac{xz}{r^2}\text{sin}^2 f,\\
S_{0}^z &= \text{cos}2f +  \frac{2z^2}{r^2}\text{sin}^2 f.
\end{aligned}
\end{equation} 
This is the simplest ansatz of a hopfion with axial symmetry about $z$-axis.
In this configuration, as shown in Fig.~\ref{fig:schematic}a, all the iso-spin contours with $S_z = 0$ form a torus surface.
Since $Q_H$ is geometrically interpreted as the linking number~\cite{link_hopf}, we show in the inset of Fig.~\ref{fig:schematic}a (upper right corner) iso-spin contours of ${\bf S}=\pm \hat{x}$, which are indeed linked.
This confirms the nontrivial topology of the spin texture under investigation.
Fig.~\ref{fig:schematic}b and c show the cross-sectional view of the spin textures at $xy$ and $yz$ plane, respectively.
A $Q_H = 1$ hopfion can also be understood as a $2\pi$ twisted skyrmion string with its two ends glued together.
Therefore, a pair of skyrmion and antiskyrmion shows up at the $y$$z$ cross-section as shown in Fig.~\ref{fig:schematic}c.
%
%Such pair is closely related to the hopfion dynamics which will be discussed later.

%
As a topological soliton, hopfion has particle-like translational dynamics.
3D anisotropic nature of the configuration also allows the rotation of hopfion.
We can capture the essential dynamics of hopfion by analyzing the collective coordinates of both translational and rotational motion.
The spin configuration of hopfion at position ${\bf r} = (x, y, z)$ and time $t$ can be expressed as ${\textbf S}(\textbf{r},t) = {\textbf S}_0 (\hat{O}(\textbf{r} - \textbf{R}))$, where {\bf{R}} = ({\it X, Y, Z}) characterizes the displacement, and $\hat{O}$ is the rotation operator.
At infinitesimal rotation, $\hat{O} \approx 1 - \bm{{\Theta}} \cdot  \hat{\textbf{L}}$, where $\hat{L_{i}} = \varepsilon_{ijk} r_j \partial_{k}$ is the angular momentum operator and $\bm{\Theta} = ({\it \Theta}_x, {\it \Theta}_y, {\it \Theta}_z)$ is the rotation angle of hopfion around different axes.

The dynamics of localized spins is in general determined by the spin Berry phase term of the Lagrangian~\cite{auerbach_interacting_1994, tatara_microscopic_2008, zang_dynamics_2011}
\begin{equation} 
\label{Eq:Berry phase}
%\begin{aligned}
L_{BP} = \int (1 - \text{cos}\theta) \dot{\phi} dV,
%\end{aligned}
\end{equation}
where $\theta$ and $\phi$ are the polar and azimuthal angle of the localized spin $\textbf{S}$ with unit length.
By integrating out the spin configuration, the variation of the spin Berry phase term $\delta L_{BP} = \int \textbf{S} \cdot \delta \textbf{S}  \times \dot{\textbf{S}} dV$ can be written in terms of the slow-varying collective coordinate as
\begin{equation} 
\label{Eq:Berry_dynamics}
%\begin{aligned}
\delta L_{BP} = D( \Theta_x \dot{Y} - \Theta_y \dot{X}) + I \Theta_y \dot{\Theta}_x,
%\end{aligned}
\end{equation}
where $D = -\int  \textbf{S}_0 \cdot (z \partial_x \textbf{S}_0 - x \partial_z \textbf{S}_0) \times \partial_x \textbf{S}_0 dV$
%$=\int x \textbf{S}_0 \cdot \partial_z \textbf{S}_0 \times \partial_x \textbf{S}_0 dV = \int y \textbf{S}_0 \cdot \partial_z \textbf{S}_0 \times \partial_y \textbf{S}_0 dV  $, 
and $I = \int \textbf{S}_0 \cdot (z \partial_x - x \partial_z) \textbf{S}_0 \times (y \partial_z - z \partial_y)\textbf{S}_0 dV $.
In Eq.~\ref{Eq:Berry_dynamics}, all other terms drop out due to parity of the spin configuration.
It clearly shows the rotations about $x-$ and $y-$ axes are canonical conjugate to each other. 
Through the entanglement between the displacement and rotation, translations along $x-$ and $y-$ directions are intertwined as well.
The longitudinal motion of hopfion is thus accompanied by transverse displacement and complex rotations.

It is noted that the $z-$axis related displacement ($Z$) and rotation ($\it \Theta_z$) are missing in Eq.~\ref{Eq:Berry_dynamics} due to the symmetry of hopfion configuration.
To capture these dynamics, it is necessary to include the auxiliary dilation of the hopfion configuration $\textbf{S}(\textbf{r},t) = \textbf{S}_0(\lambda \textbf{r}) $, where $\lambda$ is a time-dependent dilation factor and at equilibrium $\lambda = 1$.
The variation with respect to $\lambda$ then contributes an additional term to the spin Berry phase
\begin{equation} 
\label{Eq:Berry_dynamics_Z}
%\begin{aligned}
\delta L^z_{BP} = \it (\Omega \dot{Z} + \Gamma \dot{\Theta}_z) \delta \lambda,
%\end{aligned}
\end{equation}
where ${\it \Omega} = \int \textbf{S}_0 \cdot (\textbf{r} \cdot \partial_{\textbf{r}} \textbf{S}_0 \times \partial_z\textbf{S}_0 ) dV$ 
and ${\it \Gamma} = \int \textbf{S}_0 \cdot (\textbf{r} \cdot \partial_{\textbf{r}} \textbf{S}_0 \times (x \partial_y - y \partial_x)\textbf{S}_0 ) dV$.
This additional term shows the dilation is conjugated to both the displacement and rotation about $z-$axis.
Equation of motion taken from variation of $\lambda$ leads to the simultaneous translation and rotation.
It should be noticed that the dilation is not a collective coordinate since an energy change is associated with a dilation of the configuration.
Nevertheless, it plays an important role in correctly determining the corresponding hopfion dynamics.
%
%So it is important to include this dilation term in order to fully analyze the hopfion dynamics.
%
Eq.~\ref{Eq:Berry_dynamics} and Eq.~\ref{Eq:Berry_dynamics_Z} illustrate that the hopfion moves in a phase space where displacement, rotation and dilation are all entangled to each other.
%
%Through analyzing the spin Berry phase, a hopfion is expected to exhibit exotic dynamics rather than simply independent translational and rotational motions.

\begin{figure}
\begin{center}
\includegraphics[width=3.5in,height=2.6in]{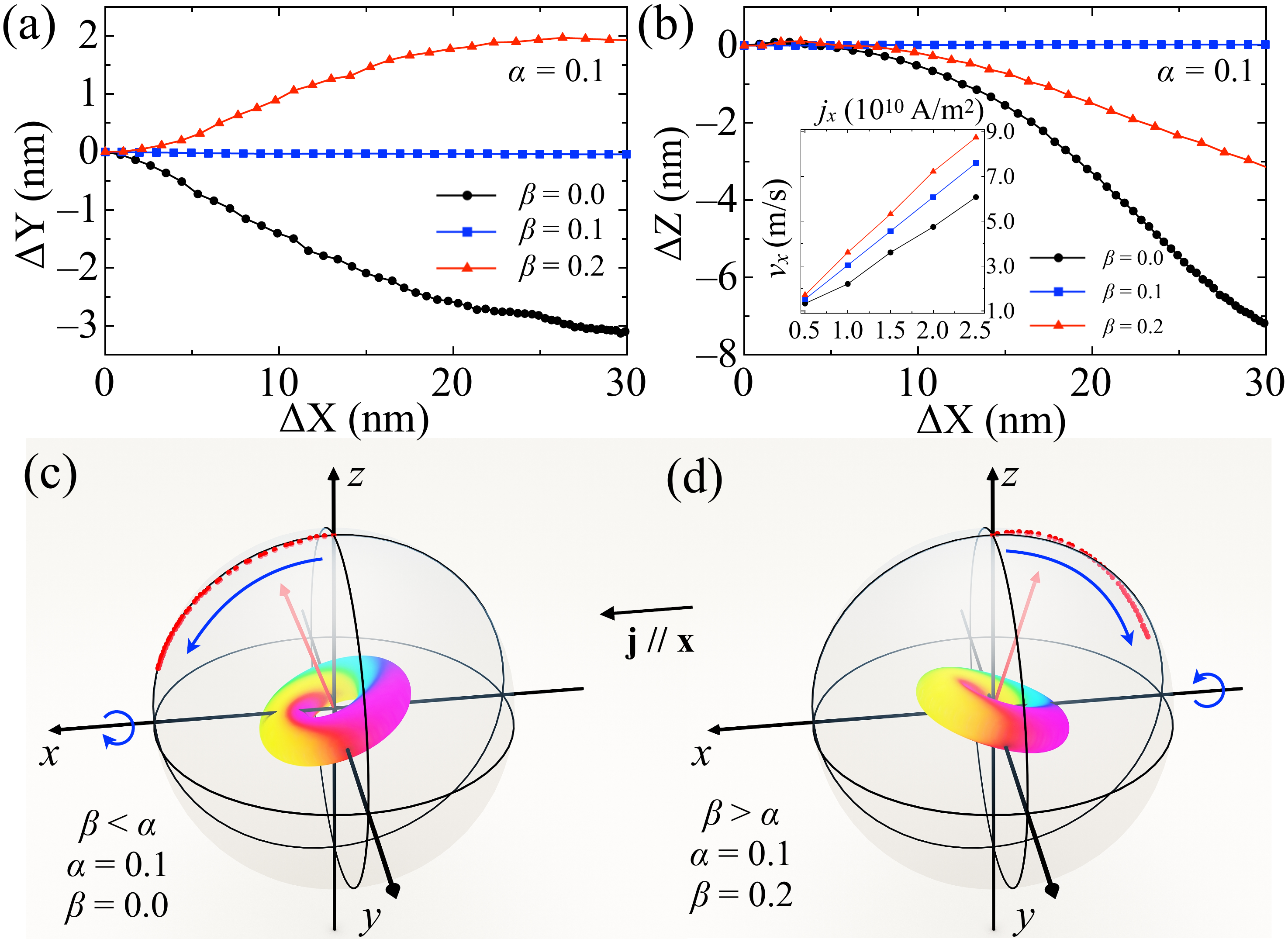}
\caption{Hopfion dynamics in the presence of an in-plane current applied in $x-$direction ($j_x=0.5 \times 10^{10} \text{A}\text m^{-2}$).
(a) and (b) show the displacements of hopfion center in $y$ and $z$ direction ($\Delta$Y and $\Delta$Z) versus the displacement in $x$ direction ($\Delta$X) for different values of $\beta$.
Inset: Current density dependence of the longitudinal velocity ($v_x$) for different values of $\beta$.
(c), (d) Rotational motion of hopfion with $\beta<\alpha$ (c) and $\beta>\alpha$ (d).
The red arrows represent the normal vector of the hopfion's midplane.
The red dots are the corresponding angles of the torus midplane projected onto a unit sphere at different simulation time.
The blue arrows indicate the direction of rotation.
}
\label{fig:in_plane}
\end{center}
\end{figure}

{\em Current-driven hopfion dynamics}---
To validate our analysis, numerical simulations were performed in order to precisely capture the hopfion dynamics.
We employ a frustrated Heisenberg Hamiltonian $\mathcal{H} =  - \sum\limits_{<i,j>} J_{ij}  \textbf{S}_i \cdot  \textbf{S}_j$, in which the summation of the exchange interaction is extended up to fourth nearest neighbor.
Here we choose the following parameters $J_1=1$, $J_2=-0.164$, $J_3=0$, $J_4=-0.082$, where the sub-indices represent the orders of nearest neighbors and all the energy terms are normalized to the value of $J_1$, the nearest neighbor exchange~\cite{supp}. 
We choose $\sin f=2r\lambda/(r^2+\lambda^2)$ as the initial state where $\lambda=1$.
A stable hopfion configuration (Fig.~\ref{fig:schematic}) is obtained by a direct energy minimization.
%
%Actually taking $\lambda$ in a reasonable range of values around $1$, we always get the same relaxed stable configuration.
%
Symmetry of this stable configuration is the same as the prototype showing by Eq.~\ref{Eq:ansatz}.

The magnetization dynamics were calculated by solving the Landau-Lifshitz-Gilbert (LLG) equation with the STT terms:
\begin{equation} 
\label{Eq:LLG}
%\begin{aligned}
\begin{split}
\frac{d\textbf{S}}{dt} = &-\gamma \textbf{S} \times \textbf{B}^{\text{eff}} + \frac{\alpha}{S} \textbf{S} \times \frac{d\textbf{S}}{dt} \\
&  + \frac{P a^3}{2eS} (\textbf{j} \cdot \nabla) \textbf{S}- \frac{P a^3 \beta}{2eS^2}\textbf{S} \times (\textbf{j} \cdot \nabla) \textbf{S}.
\end{split}
%\end{aligned}
\end{equation}
Here $\gamma$ is the gyromagnetic ratio, $\alpha$ is the damping constant, $P$ is the spin polarization, $a$ is the lattice constant and $\textbf j$ is the current density.
$\textbf{B}^{\text{eff}} =  - \frac{1}{\mu_B S} \frac{\partial \mathcal{H}}{\partial S}$ is the effective magnetic field and $S$ is the spin length, which is fixed to be 1 here for simplicity.
The last two terms in Eq.~\ref{Eq:LLG} describe the STT effect induced by an applied current $\textbf{j}$ and $\beta$ quantifies the non-adiabatic STT effect.

We begin with the current applied in the $xy$ plane.
The simulation results for a current applied along the $x-$axis are summarized in Fig.\ref{fig:in_plane}.
The hopfion dynamics can be better illustrated by using its center position and normal vector of the torus midplane, as shown in Fig.~\ref{fig:schematic}a.
At the initial state, the center position is located at the origin, the midplane lies in the $xy$ plane and its normal vector is aligned with $z-$axis. 
In the case with $\beta = 0$ and $\alpha=0.1$ ($\beta<\alpha$), two transverse motions ($\Delta Y$ and $\Delta Z$) are associated with a longitudinal motion ($\Delta X$) along the current direction (Fig.~\ref{fig:in_plane}a and b).
Meanwhile, Fig.~\ref{fig:in_plane}c shows the evolution (red dots) of the directional vector normal to the midplane (red arrow), which describes the rotation of the hopfion.
%
%It is found that the rotation around $y-$axis ($\it \Theta_y$) is much larger than that around $x-$axis ($\it \Theta_x$).
%
%So the midplane tends to rotate towards the plane ($yz$ plane in this case) perpendicular to the applied current direction ($x-$axis).
%

%

More interestingly, the non-adiabatic $\beta$-term significantly affects the hopfion dynamics.
In the case with $\beta = 0.2$ and $\alpha=0.1$ ($\beta > \alpha$), sign of $\Delta Y$ is reversed while that of $\Delta Z$ is unchanged compared to $\beta<\alpha$ case (Fig.~\ref{fig:in_plane}a and b).
In contrast, for the rotational motion, the sign of both $\it \Theta_x$ and $\it \Theta_y$ are reversed as shown in Fig.~\ref{fig:in_plane}d.
However, once $\beta  = \alpha$, all transverse motions and rotations are suppressed, and the hopfion moves straight ahead along the current direction.
For more comprehensive details of the hopfion dynamics, see movies in the Supplemental Material.

To further understand the dynamics, we derive the equations of motion for hopfion in the presence of STT effect.
A conventional approach proposed by Thiele is to first apply the operator $\partial \textbf{S}_0/\partial r_i \cdot (\textbf{S}_0 \times)$ on both sides of the LLG equation, so that the velocity on the left hand side equals to the force density on the right~\cite{thiele_steady_state_1973,tretiakov_dynamics_2008}.
However, such approach describes the translational motion only.
Notice that the term $\partial \textbf{S}_0/\partial r_i$ can be understood as the momentum operator acting on the spins.
Therefore, we can generalize the Thiele's approach by applying the operator $\hat{L} \textbf{S}_0 \cdot (\textbf{S}_0 \times)$ on both sides of the LLG equation where $\hat{L}$ is the angular momentum operator introduced in the spin Berry phase part.
In this way, we can get additional terms relating the angular velocity to the torque density.
The full set of equations of motion are summarized as (for details of the derivations, see Supplementary Materials~\cite{supp}):
\begin{equation} 
\label{Eq:IP}
\begin{aligned}
%\begin{split}
 D \dot{ \it \Theta_x} + \alpha K_{RR} \dot{Y}  + \alpha K_{R \it \Theta} \dot{ \it \Theta_y} = &\beta K_{RR} \xi j_y,
\\
-D \dot {\it \Theta_y} +\alpha K_{RR} \dot{X} + \alpha K_{R \it \Theta} \dot{\it \Theta_x} = &\beta K_{RR} \xi j_x ,
 \\
D \dot{X}  - I \dot{\it \Theta_x} +\alpha K_{R\it \Theta} \dot{Y}+ \alpha K_{\it \Theta \Theta} \dot{\it \Theta_y} =  &\beta K_{R \it \Theta} \xi j_y \\&+D\xi j_x,
\\
- D\dot{Y} +I \dot{\Theta_y} + \alpha K_{R \it \Theta} \dot{X} +\alpha K_{\it \Theta \Theta} \dot{\it \Theta_x} = &\beta K_{R \it \Theta} \xi j_x \\&- D \xi j_y, 
%\end{split}
\end{aligned}
\end{equation}
with $\xi = \frac{P a^3}{2e}$. 
$K_{RR} = \int \partial_x \textbf{S}_0 \cdot \partial_x \textbf{S}_0  dV$, 
%= \int \partial_y \textbf{S}_0 \cdot \partial_y \textbf{S}_0  dV$
$K_{R \it \Theta} = \int \partial_x \textbf{S}_0 \cdot (y \partial_z - z\partial_y) \textbf{S}_0dV$, 
%=  \int \partial_y \textbf{S}_0 \cdot (z \partial_z - x\partial_y) \textbf{S}_0dV$
and $K_{\it \Theta \Theta} =  \int [(y \partial_z - z\partial_y) \textbf{S}_0]^2 dV$
% = \int [(y \partial_z - z\partial_y) \textbf{S}_0]^2 dV$ 
are components of the dissipative tensor.  
The non-dissipative terms, namely the terms without $\alpha$ on the left hand side of each equation, are consistent with the Berry phase analysis, indicating such general approach is a proper method for handling hopfion dynamics.
By solving Eq.~\ref{Eq:IP} for a current applied along $x-$axis ($j_x$), we have $\dot{X} \sim j_x$, $\dot{Y} \sim (\alpha-\beta) j_x$, $\dot{{\it \Theta}_x}\sim (\alpha-\beta) j_x$, and $\dot{{\it \Theta}_y} \sim (\alpha-\beta) j_x$.
$\dot{Y}$,  ${{\it \Theta}_x}$, and ${{\it \Theta}_y}$ all depend on $(\alpha - \beta)$ so that their signs depend on the ratio between $\beta$ and $\alpha$.
Once $\alpha = \beta$, only $\dot{X}$ has non-zero value and only a translational motion along the current direction is allowed.
Finally, the longitudinal velocity $v_x=\dot{X}$ is linearly proportional to the current density $j_x$.
All these results are consistent with the hopfion dynamics shown in Fig.~\ref{fig:in_plane}.

While Eqs.~\ref{Eq:IP} can capture the hopfion dynamics with current in the midplane, the dynamics associated to the current component perpendicular to the midplane is completely missing.
To imitate the discussion of spin Berry phase (Eq.~\ref{Eq:Berry_dynamics_Z}), an auxiliary dilation term is included in order to fully understand the hopfion dynamics.
Under the small dilation approximation ($\lambda \sim 1$), the processional term related to ${\mathbf B}^{\text {eff}}$ can be still neglected~\cite{supp}.
In addition to the linear momentum and angular momentum approaches applied before, we can apply $(\textbf{r}\cdot\partial_{\textbf{r}}) \textbf{S}_0 \cdot (\textbf{S}_0 \times)$ on both sides of the LLG equation and then get the equations of motion along normal direction to the torus midplane, 
\begin{equation} 
\label{Eq:OOP}
\begin{aligned}
%\begin{split}
\dot{Z} &=  \frac{K_1 {\it \Omega} - K_2 {\it  \Gamma} (\beta/\alpha)}{K_1 {\it \Omega} - K_2 {\it  \Gamma} } \xi j_z, 
\\
\dot{\Theta_z} &= -\frac{K_2 {\it  \Omega}}{K_1 {\it \Omega} - K_2 {\it  \Gamma} }(1-\beta / \alpha)\xi j_z, 
\\
\dot{\lambda_z} &= -\alpha \frac{{\it  \Omega \Lambda}}{K_1 {\it \Omega} - K_2 {\it  \Gamma} }(1-\beta / \alpha)\xi j_z, 
%\end{split}
\end{aligned}
\end{equation}
with $\Lambda = K_{RR}^z K_{\it \Theta \Theta}^z - (K_{R\it \Theta}^z)^2$, 
$K_1 = \it \Omega K_{\it \Theta \Theta}^z - \Lambda K_{R\it \Theta}^z$,
and $K_2 = \it \Omega K_{R\it \Theta}^z - \Lambda K_{RR}^z$. 
The parameters $K_{RR}^z$, $K_{R\it \Theta}^z $, and $K_{\it \Theta \Theta}^z$  are defined as: 
$K_{RR}^z = \int (\partial_z \textbf{S}_0)^2 dV$,
$K_{R\it \Theta}^z = \int \partial_z \textbf{S}_0 \cdot (x\partial_y - y \partial_x) \textbf{S}_0dV$,
and $K_{\it \Theta \Theta}^z = \int [(x\partial_y - y \partial_x) \textbf{S}_0]^2 dV$.
It needs to be emphasized that in Eq.~\ref{Eq:IP} and Eq.~\ref{Eq:OOP}, the current direction is relative to the midplane of the hopfion's torus configuration.
During the hopfion dynamics, the coordinate must be co-rotating as well.
%The hopfion dynamics can be readily captured by combining these equations of motion.
%

\begin{figure}
\begin{center}
\includegraphics[width=3.5in,height=2.8in]{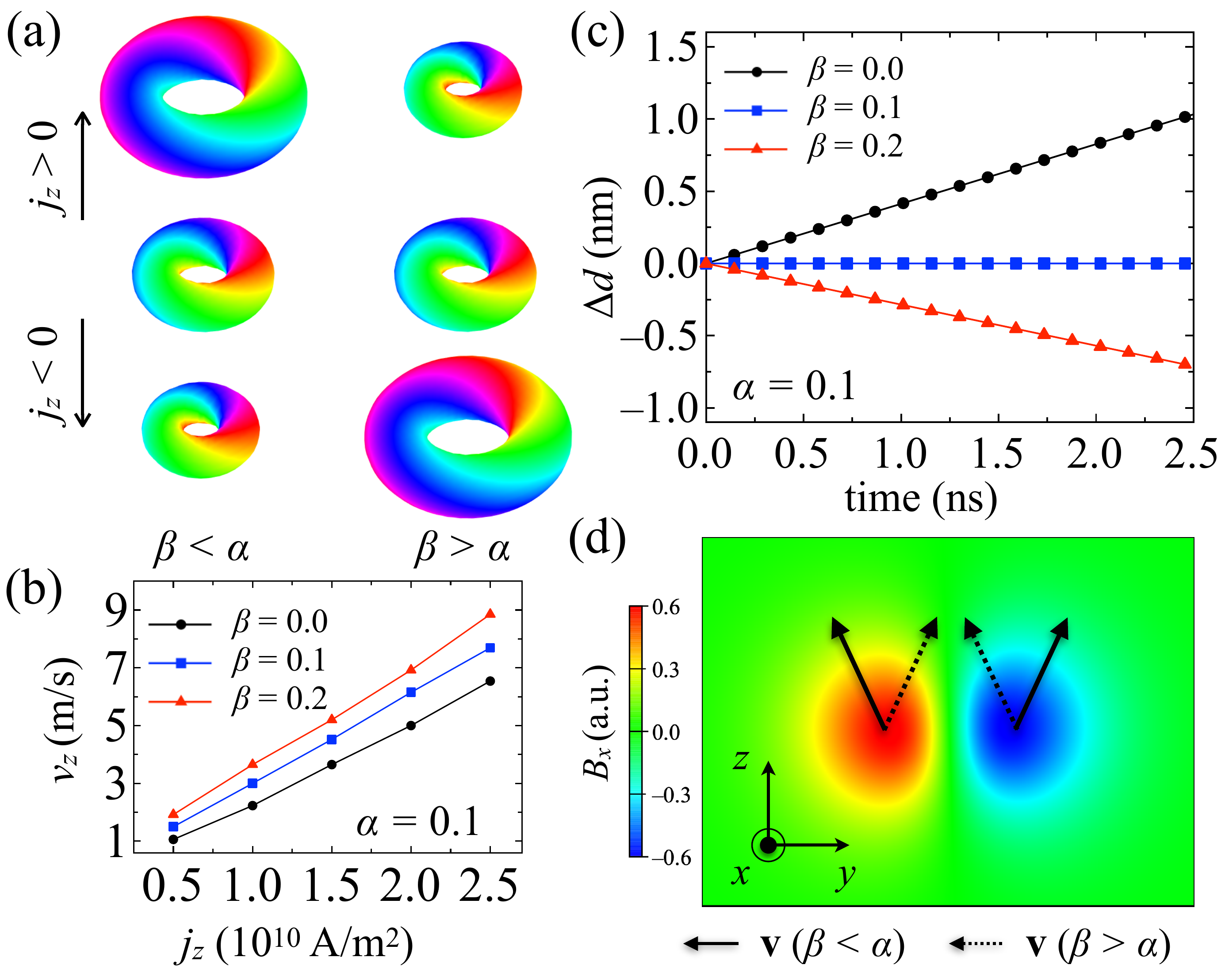}
\caption{(a) Hopfion dynamics under out-of-plane applied current ($j_z$).
A translational motion along the current direction is associated with a dilation depending on the ratio $\beta/\alpha$.
(b) Current density dependence of the hopfion velocity ($v_z$) for different values of $\beta$.
(c) Diameter change of hopfion during its translational motion ($j_z = 0.5 \times 10^{10} \text{A}\text m^{-2}$).  
(d) Calculated $B_x$ based on the spin texture shown in Fig.~\ref{fig:schematic}c. 
The arrows represent the velocities of skyrmion and antiskyrmion under an applied current along positive $z$ direction for $\beta<\alpha$ (solid) and $\beta>\alpha$ (dashed).
}
\label{fig:out_of_plane}
\end{center}
\end{figure}

Combining these equations of motion, the hopfion dynamics shown in Fig.~\ref{fig:in_plane} can be readily understood in the following way.
The current $j_x$ first induces an entangled dynamics including the longitudinal motion ($\Delta X$), transverse motion ($\Delta Y$) and rotations (${\it \Theta}_x$ and ${\it \Theta}_y$).
As the midplane of hopfion starts to deviate from the $xy$ plane, the current can be decomposed into two components, one in the midplane ($j_{\parallel}$) and one normal to the midplane ($j_{z}$).
While the former component is still responsible for the entangled dynamics mentioned above, the hopfion motion $\Delta Z$ along normal direction starts to develop according to Eq.~\ref{Eq:OOP}.
%
%The midplane is initially overlapped with the $xy$ plane, resulting in a small amplitude of $j_{z}$, which further explains the small value of $\Delta Z$ at the beginning (Fig.~\ref{fig:in_plane}b).

To examine the dynamics in the normal direction, we study the hopfion dynamics under $j_z$.
The corresponding simulation results are summarized in Fig.~\ref{fig:out_of_plane}a-c.
The current induces a translational motion of hopfion along its direction in combination with a dilation and rotation about $z-$axis.
The dilation type is determined by the ratio between $\beta$ and $\alpha$ (Fig.~\ref{fig:out_of_plane}c).
When $\beta<\alpha$, the hopfion is compressed (expanded) by a negative (positive) current, and the case is reversed for $\beta>\alpha$.
%
%A rotation around $z-$axis (${\it \Theta}_z$) is also present in both cases.
%
While for $\beta=\alpha$, both dilation and rotation are absent.
It is worth mentioning that the expansion and compression of hopfion are not quite symmetric since there is an energy barrier to prevent further compression of hopfion in order to maintain its topology.
Similarly to the current in plane case, the velocity of hopfion $v_z=\dot{Z}$ here is also linearly proportional to the current density (Fig.~\ref{fig:out_of_plane}b).
All these dynamics are well described by Eqs.~\ref{Eq:OOP}.

The interesting dilation of hopfion can be also understood phenomenologically in terms of the skyrmion string. 
As mentioned earlier, a hopfion can be recognized as a $2\pi$ twisted skyrmion string with its two ends glued together and thus a skyrmion-antiskyrmion pair is formed in any cross-section plane including the $z-$axis (e.g., $xz$ or $yz$ plane), similar to that shown in Fig.~\ref{fig:schematic}c.
To further illustrate the hopfion dynamics, the emergent magnetic field $B_i = \frac{1}{2} \varepsilon_{ijk} \textbf{S}\cdot (\partial_{j} \textbf{S}\times \partial_{k}\textbf{S})$ is calculated based on the hopfion configuration in Fig.~\ref{fig:schematic}c and the $B_x$ is shown in color in Fig.~\ref{fig:out_of_plane}d.
It is known that the current-driven motion of skyrmion has a transverse component, i.e., the skyrmion-Hall effect~\cite{zang_dynamics_2011, nagaosa_topological_2013, jiang_direct_2017, litzius_skyrmion_2017}.
The corresponding skyrmion-Hall angle is determined by the topological charge, or in identical terms, the emergent magnetic field of the skyrmion.
More importantly, the sign of the skyrmion-Hall angle depends the ratio between $\beta$ and $\alpha$~\cite{nagaosa_topological_2013, iwasaki_universal_2013} as shown by the arrows in Fig.~\ref{fig:out_of_plane}d.
As a result, the skyrmion-antiskyrmion pair shown in Fig.~\ref{fig:out_of_plane}d respond to a current in $z-$direction by moving towards or away from each other during their motion along $z$.
Same is true for any cross section slicing the hopfion.
When the skyrmion and antiskyrmion move towards (away from) each other, the hopfion is compressed (expanded).
The hopfion dynamics can thus also be phenomenologically understood as a collective motion of skyrmion-antiskyrmion pairs
% and it further highlights the fundamental connections between the magnetization dynamics and the fluid dynamics~\cite{moffatt_helicity_2014}.

{\em Conclusion}---
Current-driven 3D dynamics of magnetic hopfion have been studied both analytically and numerically.
The hopfion exhibits rich dynamics of entangled translation, rotation and dilation.
The theory built upon spin Berry phase and generalized Thiele's approach gives out simple equations of motion reproducing numerical results.
%These different types of dynamics are entangled together through the spin Berry phase of hopfion and can be captured by the derived equations of motion.
%
%Since dilation is included in the equations of motion, new approaches beyond the collective coordinate may be needed for further hopfion studies.
%
Our phenomenological analysis also reveals the vital role of skyrmion-antiskyrmion pair in hopfion dynamics, and makes connection of soliton dynamics across dimensionality.
Since our theory is built on the collective coordinates that is independent of details of spin interactions, it suggests the universality of the reported dynamics in all existing and forthcoming hopfion models, not only in magnetism, but also in other physical systems~\cite{moffatt_energy_1990, faddeev_stable_1997, manton_topological_2004, radu_stationary_2008}.
The rich dynamics hosted by a $Q_H=1$ hopfion further foreshadows more exotic dynamics for hopfions with higher $Q_H$ and their potentials in spintronic applications~\cite{fernandez_3dimensional_2017}.

{\em Acknowledgement}---
J.Z. acknowldges the finical support by the U.S. Department of Energy (DOE), Office of Science, Basic Energy Sciences (BES) under Award No. DE-SC0020221.
Y.L and X.H. thank the finical supports from the National Key Research and Development Program of China (Grant No. 2017YFA0206200, 2016YFA0300802), the National Natural Science Foundation of China (NSFC, Grants No.51831012, 11804380).

%\bibliography{hopf}

\end{document}